\documentstyle[12pt,epsf]{article} 

\def\simg{{\ \lower-1.2pt\vbox{\hbox{\rlap{$>$}\lower6pt\vbox{\hbox{$\sim$}}}}\ }}
\def\siml{{\ \lower-1.2pt\vbox{\hbox{\rlap{$<$}\lower6pt\vbox{\hbox{$\sim$}}}}\ }} 
\def\bfnabla{\mbox{\boldmath $\nabla$}}

\def\bfsigma{\mbox{\boldmath $\sigma$}}
\def\als{\alpha_{s}}
\def\al{\alpha}
\def\lQ{\Lambda_{\rm QCD}}

\def\dsl{\,\raise.15ex\hbox{/}\mkern-13.5mu D}

\newcommand{\nn}{\nonumber}
\newcommand{\be}{\begin{equation}}
\newcommand{\ee}{\end{equation}}
\newcommand{\bea}{\begin{eqnarray}}
\newcommand{\eea}{\end{eqnarray}}

\newcommand{\Appendix}[1]%
    {%
     \section{#1}%
      }

\begin{document}\setlength{\unitlength}{1mm}

\begin{titlepage}
\begin{flushright}
\tt{UB-ECM-PF 02/09}
\end{flushright}

\vspace{1cm}
\begin{center}
\begin{Large}
{\bf Renormalization group improvement of the spectrum of
Hydrogen-like atoms with massless fermions}\\[2cm]
\end{Large} 
{\large Antonio Pineda}\footnote{pineda@ecm.ub.es}\\
{\it Dept. d'Estructura i Constituents de la Mat\`eria and IFAE,
  U. Barcelona \\ Diagonal 647, E-08028 Barcelona, Catalonia, Spain
        \\}
\end{center}

\vspace{1cm}

\begin{abstract}
We obtain the next-to-next-to-leading-log renormalization group
  improvement of the spectrum of Hydrogen-like atoms with massless
  fermions by using potential NRQED. These results can also be applied
  to the computation of the muonic Hydrogen spectrum where we are able
  to reproduce some known double logs at $O(m\als^6)$. We compare with
  other formalisms dealing with log resummation available in the
  literature. \vspace{5mm} \\ PACS numbers: 11.10.St, 11.10.Hi,
  12.38.Cy, 12.38.Bx
\end{abstract}

\end{titlepage}
\vfill
\setcounter{footnote}{0} 
\vspace{1cm}

In Ref. \cite{rgmass} (see also \cite{rgpot}), the renormalization
group (RG) improvement of the Heavy Quarkonium spectrum for the equal
mass case was obtained within the potential NRQCD (pNRQCD) formalism
\cite{pNRQCD}. This result was compared with the one of Ref. \cite{HMS}
(see also \cite{vNRQCD1,vNRQCD2}) obtained within the vNRQCD formalism
\cite{vNRQCD}. Disagreement was found. This disagreement is potentially
important as it propagates to different observables like, for example, $t$-$\bar t$
production near threshold, where it is claimed \cite{HMST} that the
resummation of logs plays an important role. For instance, the
matching coefficient of the electromagnetic current, which is a
necessary ingredient in these calculations, is different \cite{rgcurr,vNRQCD2}.
Nevertheless, for the known logs at
next-to-next-to-leading \cite{leadinglog} and next-to-next-to-next-to-leading
order \cite{curNNLL}, both calculations happen to agree with each other.

In order to try to clarify this issue, we will consider the simplified
problem of a Hydrogen-like system coupled to $n_f$ light (massless)
fermions in QED. We will then obtain the next-to-next-to-leading-log
(NNLL) RG scaling of the spectrum of this system. In principle, these results can be
applied to muonic Hydrogen. In this case, the electron is replaced by the muon, $n_f
\rightarrow 1$ and the remaining light fermion is the electron (which we
will take to be massless for simplicity or, at most, of $O(m\al^2)$,
where $m$ is the mass of the muon). In this situation, we will be able to compare, in
certain limits, with finite $O(m\al^6\ln^2)$ results already available
in the literature \cite{Pachucki}. Our results will agree with these
calculations.

The computation will closely follow the procedure of
Ref. \cite{rgmass} to which we refer for details. Here we will just
write the main formulas necessary to set up the notation and the
results.

The first step is to obtain the RG improved
matching coefficients of the NRQED \cite{NRQED} Lagrangian at one loop
and up to $O(1/m^2)$ ($m$ is the mass of the massive lepton (the muon
for the muonic hydrogen) and the mass of the nucleus is sent to
infinity in this paper). 

The NRQED Lagrangian including light fermions reads at $O(1/m^2)$ (up
to field redefinitions) \cite{NRQED,previous,BM}
\be
\label{LagNRQCD}
{\cal L}={\cal L}_{ph}+{\cal L}_l+{\cal L}_{\mu}+{\cal L}_{p}+{\cal
L}_{\mu p}
\,,
\ee
where $\mu$ is the Pauli spinor that annihilates the fermion, $N_p$
is the Pauli spinor that annihilates the nucleus, $i D_0=i\partial_0
-gA_0$, $i{\bf D}=i\bfnabla+g{\bf A}$, 
\be
\label{Lg}
{\cal L}_{ph}=-\frac{1}{4}F^{\mu\nu}F_{\mu \nu} 
 ,
\ee
\bea
\label{Ll}
{\cal L}_l&=&\sum_i \bar l_i i \dsl l_i+
c_1^{ll}\displaystyle \frac{g^2}{8m^2}\sum_{i,j}\bar{l_i}  \gamma^\mu
 l_i \ \bar{l}_j \gamma_\mu l_j +
c_2^{ll}\displaystyle \frac{g^2}{8m^2}\sum_{i,j}\bar{l_i} \gamma^\mu
\gamma_5
 l_i \ \bar{l}_j \gamma_\mu \gamma_5 l_j,
\eea
\bea
\label{Lhl}
{\cal L}_{\mu}&=&
\mu^{\dagger} \Biggl\{ i D_0
+ \, c_k{{\bf D}^2\over 2 m} + \, c_4{{\bf D}^4\over 8 m^3}
+ c_F\, g {{\bf \bfsigma \cdot B} \over 2 m}
\\ \nonumber
&& \qquad
+ c_D \, g { \left({\bf D \cdot E} - {\bf E \cdot D} \right) \over 8 m^2}
+ i c_S \, g { {\bf \bfsigma \cdot \left(D \times E -E \times D\right) }\over 8 m^2} 
\Biggr\} \mu
\\
\nn
&&
+c_1^{\mu l}\displaystyle\frac{g^2}{8m^2}\sum_i\mu^{\dagger}
  \mu \ \bar{l}_i\gamma_0 l_i
+c_2^{\mu l}\displaystyle\frac{g^2}{8m^2}\sum_i\mu^{\dagger}\gamma^\mu\gamma_5
\mu 
  \ \bar{l}_i\gamma_\mu\gamma_5 l_i,
\eea
\be
{\cal L}_{p}=N_p^{\dagger}iD^0_pN_p
\,,
\ee
where $iD^0_p=i\partial_0+gZA_0$ and
\be
{\cal L}_{\mu p} =
  {d_{s} \over m^2} \mu^{\dag} \mu N_p^{\dag} N_p
+ {d_{v} \over m^2} \mu^{\dag} {\bfsigma} \mu
                         N_p^{\dag} {\bfsigma} N_p
\,.
\ee
We have also included the ${\bf D}^4/m^3$ term above since it will be necessary
in the evaluation of the spectrum once the power counting
is established. Moreover, we will consider that the kinetic term
matching coefficients are protected by reparameterization invariance
($c_k=c_4=1$) \cite{RI}, however, we will often keep them explicit for
tracking purposes. 

By definition, NRQED has an ultraviolet cutoff $\nu_{\rm
NR}=\{\nu_p,\nu_{s}\}$ satisfying $mv \ll \nu_{\rm NR} \ll m$. $\nu_p$ is
the ultraviolet (UV) cut-off of the relative three-momentum of
the heavy fermion and antifermion. $\nu_s$ is the UV cut-off of the
three-momentum of the photons and light fermions. The derivation of the
scale dependence of the matching coefficients with respect the UV
cutoffs of the theory goes identical to the one in
Ref. \cite{rgmass}. In particular the fact that no dependence of $\nu_p$
appears at this order.  
In principle, the running of $c^{ll}$ and $c^{\mu l}$ could be deduced
from the results of \cite{BM,previous} by taking care of the changes
of the color structure. Since we are only interested in the
computation of the spectrum at NNLL accuracy, their contribution will
vanish at this order as far as the spectrum is concerned ($c_1^{\mu l}$
appears in the equation of $c_D$ 
but the running of $c_1^{\mu l}$ is zero at LL accuracy). 
Therefore, the relevant RG equations in our case read  
\be
\nu_s {d\over d\nu_s}c_D=-{\alpha\over  \pi}
\left(
{8 \over 3}c_k^2+{\beta_0 \over 2}c_1^{\mu l}
\right)
\label{RGeqhcoulomb}
\ee
and zero otherwise.

By taking the matching conditions at the scale $m$: $c_k=c_F=c_s=c_D=1$
and $\{d\}=0$, 
we can obtain the solution of the RG equations. We only
explicitely display those which will be necessary later on (we
define $z=\left[{\al(\nu_s) \over \al(m)}\right]^{1 \over
\beta_0}\simeq 1 -1/(2\pi)\al(\nu_s)\ln ({\nu_s \over m})$,
$\beta_0=-{4 \over 3}T_Fn_f$ with $T_F=1$)
\bea
c_F(\nu_s)&=&1
\,,
\nn\\
c_S(\nu_s)&=&1
\,,
\nn\\
c_D(\nu_s)&=&
1+{16 \over 3}\ln{z}
\,,
\nn\\
d_{s}(\nu_s)&=&0
\,,
\nn\\
d_{v}(\nu_s)&=&0
\label{RGeqhs}
\,.
\eea

\medskip

The above results are a necessary step towards the RG improvement of
pNRQED with the matter content described above, which we consider in
what follows.
pNRQCD is defined by the cut-off
$\nu_{\rm pNR}=\{\nu_p,\nu_{us}\}$, where $\nu_p$ is
the cut-off of the relative three-momentum of the heavy fermions and is
such that $mv \ll \nu_p \ll m$ and $\nu_{us}$ is the cut-off of the
three-momentum of the photons and light fermions with $mv^2 \ll \nu_{us} \ll mv$.
 
The pNRQED Lagrangian reads as follows ($iD^0_S=i\partial_0+g(Z-1)A_0$):
\bea
&&L_{\rm pNRQED} =
\label{pnrqcdph}
\int d^3{\bf x} d^3{\bf X} S^{\dagger}({\bf x}, {\bf X}, t)
                \Biggl\{
iD^0_S - c_k{ {\bf p}^2 \over 2m} + c_4{ {\bf p}^4 \over 8m^3}+ 
\\
&&
\nonumber
- V^{(0)}- {V^{(1)} \over m}- {V^{(2)} \over m^2}+ g V_A{\bf x} \cdot {\bf E} ({\bf X},t)
\Biggr\}
S ({\bf x}, {\bf X}, t)- \int d^3{\bf X} {1\over 4} F_{\mu \nu} F^{\mu \nu}
\,,
\eea
where ${\bf x}$ and ${\bf X}$, and ${\bf p}$ and ${\bf P}$ are the
relative and center of mass coordinate and momentum respectively. All
the gauge fields in Eq. (\ref{pnrqcdph}) are functions of the
center-of-mass coordinate and the time $t$ only. We have explicitly
written only the terms relevant to the analysis at the NNLL.

We now display the structure of the matching potentials $V^{(0)}$,
$V^{(1)}$ and $V^{(2)}$, which are the relevant ones to our
analysis. At order $1/m^0$, we have the static potential:
\begin{equation}
V^{(0)} \equiv  - Z {\alpha_{V} \over r}.
\label{defpot0s}
\end{equation}
In principle, at order $1/m$, we may have a potential scaling as
${V^{(1)} \over m} \sim {1 \over mr^2}$.  Nevertheless, it vanishes at
the order we are working. It would give, at most, $O(m\al^6)$
corrections to the spectrum in a finite order calculation and the
running equations would not mix with it. Therefore, for the purposes
of this paper, we approximate
\be
{V^{(1)} \over m} \simeq 0.
\ee
At order $1/m^2$, to the accuracy we aim at, $V^{(2)}$ has the structure  
\be
{V^{(2)} \over m^2} = 
{\pi Z D^{(2)}_{d} \over m^2}\delta^{(3)}({\bf r})
+ { 3 Z D^{(2)}_{LS} \over 2 m^2}{1 \over r^3}{\bf L}_1 \cdot {\bf S}_1
,
\label{V2}
\ee
where ${\bf S}_1 = \bfsigma_1/2$. In principle, one may consider more
structures for the $1/m^2$ potential but, since they will not
contribute at the accuracy we aim and in order to focus the problem as
much as possible, we will set them to zero in what follows, as we have
done for the $1/m$ potential.

The coefficients, ${\tilde V}=\{\alpha_{V_s}$, $D_s$, ...$\}$ contain
some $\ln r$ dependence once higher order corrections to their leading
(non-vanishing) values are taken into account.  In particular, we will
have expressions like $\delta^{(3)}({\bf r})\ln^n r$. This is not a
well-defined distribution and should be understood as the Fourier
transform of $\ln^n 1/k$.  Nevertheless, in order to use the same
notation for all the matching coefficients, and since it will be
sufficient for the purposes of this paper, namely to resum the leading logs, we
will use the expression $\delta^{(3)}({\bf r})\ln^n r$, although it
should always be understood in the sense given above.

By studying the UV behavior of pNRQED
it is possible to obtain the scale dependence of the coefficients of
the potentials ${\tilde V}$. The discussion closely follows the one of
Ref. \cite{rgmass} to which we refer for details. Here we just mention
the main points. The potentials have the following structure:
\be
{\tilde V}(d(\nu_p,\nu_s,m),c(\nu_s,m),\nu_s,\nu_{us},r)={\tilde
V}(\nu_p,m,\nu_{us},r) \equiv {\tilde V}(\nu_p,\nu_{us})
\,.
\ee
In particular, 
\be
\nu_s {d\over d\nu_s}{\tilde V}=0.
\label{nus}
\ee 
Moreover, at the
accuracy we aim, we also get
\be
\label{nup1}
\nu_p {d\over d\nu_p}{\tilde V} =0.  
\ee 
Therefore, we obtain
\be
{\tilde V}(\nu_p,\nu_{us})
\simeq {\tilde V}(\nu_{us})  
\ee 
and we only have to
compute the $\nu_{us}$ scale dependence.

\medskip

The $\nu_{us}$-scale dependence could be obtained along the same lines
as in Ref. \cite{rgmass}. We obtain in this specific case:
\bea
\nu_{us} {d\over d\nu_{us}}\alpha_{s}&=&-\beta_0 {\al^2 \over 2\pi}
\,,
\label{RGeqm}
\\
\nn
\nu_{us} {d\over d\nu_{us}}D_{d}^{(2)}&=& 
-{4 \over 3}{\alpha(\nu_{us})\over
  \pi}V_A^2c_k^2 \al(r^{-1})
\,,
\eea
and zero for the other potentials.

Eqs. (\ref{nus}), (\ref{nup1}) and (\ref{RGeqm}) provide
the complete set of RG equations at the desired order. By using
Eqs. (\ref{nus}) and (\ref{nup1}), we obtain
\be
{\tilde V}={\tilde V}(d(1/r,m),c(1/r,m),\nu_s=1/r,\nu_{us},r)
\,.
\ee
We now need the initial condition in order to solve the US RG equations,
i.e. the matching conditions. We fix the initial point at
$\nu_{us}=1/r$. In summary, we need to know the static
potential with $O(\al^3)$ accuracy, the $1/m$ potential with
$O(\al^2)$ accuracy, the $1/m^2$ potentials with $O(\al)$ accuracy and $V_A$ with $O(1)$
accuracy at $\nu_{us}=1/r$. For the non-vanishing potentials, they read  
\bea
{\alpha}_{V}(r^{-1}) &=&\alpha(r^{-1})
\left\{1+\left(a_1+ 2 {\gamma_E \beta_0}\right) {\alpha(r^{-1}) \over 4\pi}\right.
\nonumber\\
&&
\left.
+\left[\gamma_E\left(4 a_1\beta_0+ 2{\beta_1}\right)+\left( {\pi^2 \over 3}+4 
\gamma_E^2\right) 
{\beta_0^2}+a_2\right] {\alpha^2(r^{-1}) \over 16\,\pi^2}
 \right\},
\label{newpot0i}\nn\\ 
D^{(2)}_{d}(r^{-1})&=& \alpha(r^{-1}){c_D(r^{-1}) \over 2}
\,,
\label{Dd2i}\nn\\ 
D^{(2)}_{LS,s}(r^{-1})&=& {\alpha(r^{-1}) \over 3}c_S(r^{-1}), 
\label{DLs2i}\nn\\ 
V_A(r^{-1})&=&1
\label{vai},
\eea
where $\beta_1=-4T_Fn_f$ and the values of $a_1$ and $a_2$ can be
easily obtained from the QCD results \cite{FSP} by taking $C_f
\rightarrow 1$, $C_A \rightarrow 0$ and $T_F \rightarrow 1$.

We now have all the necessary ingredients to solve the RG
equations. The RG improved  potentials read: 
\bea
{\alpha}_{V}(\nu_{us}) &=&{\alpha}_{V}(r^{-1}),
\label{newpot0}
\nn
\\ 
D^{(2)}_{d}(\nu_{us})&=& D^{(2)}_{d}(r^{-1})
-
{8\over 3\beta_0}\al(r^{-1})\log\left(
\alpha(r^{-1})\over \alpha(\nu_{us}) \right)
=
{\al(r^{-1}) \over 2}
\left(
1
-
{16\over 3\beta_0}\log\left(
\alpha(m)\over \alpha(\nu_{us}) \right)
\right)
,
\label{Dd2}
\nn
\\ 
D^{(2)}_{LS}(\nu_{us})&=& D^{(2)}_{LS}(r^{-1}).
\label{DLs2}
\eea
This completes the RG evaluation of the pNRQED Lagrangian at NNLL.

\medskip

With the above results we can obtain the energy with NNLL
accuracy. The discussion goes similar to the one in
Ref. \cite{rgmass}. All the large logs can be obtained from the
potential terms. Once the potentials are introduced in the
Schr\"odinger equation, the $\ln^n(1/r)$ terms produce $\ln^n(m\al)$
terms plus subleading contributions ($\ln^{n-1}(m\al)$,
$\cdots$) within the LL resummation counting. The expectation value of
the potential terms is $\nu_{us}$-scale dependent. This scale
dependence is cancelled by the ultraviolet scale dependence of
ultrasoft loops. The typical scales in these integrals is of the order
$m\al^2$. Therefore, the logs of the ultrasoft loops get minimized by
setting $\nu_{us}\sim m\al^2$ and all the large logs get encoded in
the potential contributions. Finally, one obtains the following
correction to the NNLO energy expression:
\be
 \delta E_{n,l,j}^{\rm pot}(\nu_{us}) = E_n \al^2  
{ Z^2\delta_{l0} \over 3 n}
\left(
-{16\over \beta_0}
\log\left(
\alpha(\nu_{us})\over \alpha \right) 
-3(c_D-1)
\right)
\label{energy1} 
 \,,
\ee
where $E_n= - mZ^2\al^2/(2n^2)$ and the scale $\nu_s$ in $z$ and in the
NRQED matching coefficients has been fixed to the soft scale
$\nu_s=2a_n^{-1}$, where $a_n^{-1}={mZ\al(2a_n^{-1}) \over n}$. $\al$
is also understood at the soft scale $\nu_s=2a_n^{-1}$ unless the
scale is specified. The $\nu_{us}$-scale dependence of Eq. (\ref{energy1})
cancels against contributions from US energies. Since $m\al^2$ is the
next relevant scale, their effective role will be to replace
$\nu_{us}$ by $m\al^2$ (up to finite pieces that we are
systematically neglecting) in Eq. (\ref{energy1}). In particular, we
take $\nu_{us}=-E_n$. As expected, Eq. (\ref{energy1}) with
$\nu_{us}=-E_n$ reproduces the well known Hydrogen-like
$O(m\al^5\ln\al)$ correction but, indeed, Eq. (\ref{energy1}) gives
all the $O(m\al^4(\al\ln\al)^n)$ terms for $n \geq 1$ of the spectrum
of the Hydrogen-like systems with $n_f$ massless fermions. After
adding to Eq. (\ref{energy1}) the NNLO result with the normalization
point at the {\it same} soft scale, $\nu_s=2a_n^{-1}$, that we have
used here, the complete NNLL mass is obtained. Note that the above resummation 
of logs also correctly accounts for $\ln Z$ terms with the same accuracy. 

\medskip

We have seen that the large logs of the spectrum can be obtained from
the potential terms by setting $1/r \sim m\al$ and $\nu_{us} \sim
m\al^2$. The velocity of the non-relativistic particle is typically $v
\sim Z\al$. Therefore, it is interesting to consider the scaling of the
potentials with respect $v$ as it will help us to later compare with
vNRQED results. In practice, we will consider its scaling with respect
$\nu \equiv mv$ (therefore $\nu_{us}=\nu^2/m$) where
\be
{\tilde V}(\nu_p,m,\nu_{us},r) \simeq {\tilde V}(m,\nu_{us},r) \rightarrow  
{\tilde V}(m,\nu^2/m,1/\nu) \equiv {\tilde V} (\nu)
\,.
\ee
We can now consider its derivative with respect $\nu$. We will just focus on $D^{(2)}_d$ 
since it is the only one which has a non-trivial running. We obtain  
\be
\nu {d\over d\nu}D_{d}^{(2)}=
-{\beta_0 \over 4\pi}c_D(\nu)\al^2(\nu)
+{4 \over 3}{ \al^2(\nu) \over \pi}\ln{\al(\nu) \over \al({\nu^2 \over m})}
-{8 \over 3}{\al(\nu) \over \pi}\al({\nu^2 \over m})
\,.
\ee
It is remarkable that the above expression can be rearranged as
\be
\nu {d\over d\nu}D_{d}^{(2)}=
-{\beta_0 \over 4\pi}c_D({\nu^2 \over m})\al^2(\nu)
-{8 \over 3}{\al(\nu) \over \pi}\al({\nu^2 \over m})
\,.
\label{pNRQEDrun}
\ee

\medskip

There is an evaluation \cite{HMS} within the vNRQCD framework
\cite{vNRQCD} of the RG improved Heavy Quarkonium mass when $\lQ \ll
m\als^2$. The evaluation performed within the pNRQCD framework
\cite{rgmass} disagreed with that evaluation. It was noticed there that
the disagreement still persisted if one considered a QED-like limit
with light fermions by taking $C_f \rightarrow 1$, $C_A \rightarrow 0$
and $T_F \rightarrow 1$. Agreement was found for a QED-like limit
without light fermions by taking $C_f \rightarrow 1$, $C_A \rightarrow
0$, $n_f \rightarrow 0$, $T_F \rightarrow 1$. Some errors seem to have
been detected in the first versions of these calculations in vNRQCD
\cite{Hoangpriv} that may partially explain the difference, in
particular for the $1/m^2$ potential. In this case, agreement may
exist in the limit $C_f \rightarrow 1$, $C_A \rightarrow 0$ and $T_F
\rightarrow 1$.

For the evaluation performed in this paper, the computation of 
the spectrum for the case of Hydrogen-like atoms with
massless fermions, there exists no analogous within the 
vNRQED framework.  Nevertheless, it is
possible to guess what would be the result in that formulation by
using the rules of Ref. \cite{MSS}, which relate the anomalous
dimensions computed here with the ones that should appear in
vNRQED. For the specific case of $D_d^{(2)}$, we obtain
\be
\nu {d\over d\nu}D_{d}^{(2)}({\rm vNRQED})=\gamma_s+2\gamma_u
\,,
\label{vNRQEDrun}
\ee
where 
\be
\label{anomdim}
\gamma_s=-{\beta_0 \over 4\pi}c_D(\nu)\al^2(\nu)\,,\qquad 
\gamma_{us}=-{4 \over 3}{\al(\nu) \over \pi}\al({\nu^2 \over m})
\,.
\ee
This should be compared with the running in pNRQED obtained above. If
we do so, we find that Eqs. (\ref{vNRQEDrun}) and
(\ref{pNRQEDrun}) are different. If expanded in $\al$ they first
differ at $O(\al^2\log^2\al)$. This produces a difference in the
computation of the mass at $O(m\al^6\log^2\al)$. In order to perform
an independent check, it would be extremally important that
corrections of this order had been computed before. The closest system
to the one discussed here corresponds to the muonic hydrogen for
which, indeed, corrections to the energy at this order have been
computed by Pachucki \cite{Pachucki}. In order to compare our results
with his evaluation, we have to take the limit $n_f \rightarrow 1$. Moreover,
for the real muonic hydrogen, the mass of the light fermion (the
electron in this case) is not negligible. However, we can formally
consider the situation $m_e \sim m\al^2$ (even if for the physical
situation $m_e \sim m\al$ is closer to reality) in his and our
calculation. For the matter of comparison, in our case, this means
that, for scales of the order of $m_e$ and $m\al^2$, we can use the low
energy electromagnetic coupling $\al_{\rm em} \sim 1/137...$. This is
indeed the parameter expansion used in Pachucki's calculation.  A
closer inspection shows that the diagrams that give rise to the large
logs computed here correspond to the ones drawn in Fig. 4 in
Ref. \cite{Pachucki}. If we reexpand our result in terms of $\al_{\rm
em}=\al(\nu_{us})$, we obtain (up to the order of interest and with $v
\sim \al$)
\bea
\label{expD}
D_d^{(2)}-{\al(\nu) \over 2} 
&=&
{\al(\nu_{us}) \over 2}
\left(
1+{\beta_0 \over 2\pi}\al(\nu_{us})\ln{mv^2 \over mv}+\cdots\right)
\\
\nn
&&
\times
\left(
-{8 \over 3}{\al(\nu_{us}) \over \pi}\ln{mv^2 \over m}
-{2 \over 3}\beta_0\left({\al(\nu_{us}) \over \pi}\right)^2\ln^2{mv^2 \over m}
+\cdots\right)
\\
\nn
&\simeq&
-{4 \over 3}{\al^2(\nu_{us}) \over \pi}\ln{mv^2 \over m}
-{2\beta_0 \over 3}{\al^3(\nu_{us}) \over \pi^2}\ln{mv^2 \over mv}\ln{mv^2 \over m}
-{1 \over 3}\beta_0{\al^3(\nu_{us}) \over \pi^2}\ln^2{mv^2 \over m}
\,.
\eea
It is easy to identify the above terms (last equality) within a diagrammatic
picture. The first term is the standard Lamb-shift correction one
would find for the Hydrogen atom and corresponds to the diagrams of
Fig. 4 of Ref. \cite{Pachucki} without any bubble insertion. The
second term corresponds to the first diagram in Fig. 4 of
Ref. \cite{Pachucki}. The last term corresponds to the second diagram
in Fig. 4 of Ref. \cite{Pachucki}. Therefore, our result seems to have
the correct structure for the $O(m\al^6\ln^2)$ corrections. Let us now
go deeper in the comparison with Pachucki's results. First, we can see
that the last term of Eq. (\ref{expD}) can reproduce the analogous
Pachucki's contribution by setting $\al(\nu_{us})=\al_{\rm em}$ and
$\nu_{us}=m_e$ (this result depends on the two-loop muon form factor
first computed in Ref. \cite{BCR}). For the second term of
Eq. (\ref{expD}), the explicit comparison is a little bit more
involved. Nevertheless, it is possible to see that the first term in
Eq. (39) of Ref. \cite{Pachucki} gives the logs of the second term in
Eq. (\ref{expD}) since one can make the replacement (as far as the LL
contribution is concerned) 
\be
V_{VP} \rightarrow -{Z\al(\nu_{us})  \over r}
\left[
{\beta_0 \over 2\pi}\al(\nu_{us})\ln{mv^2 \over mv}
\right]
\ee
for $V_{VP}$, as defined in
Ref. \cite{Pachucki}\footnote{We note that for this diagram both
loops factorize. Therefore, no sign of correlation of scales appears at
this level of the computation.}. The second term in Eq. (39) of
Ref. \cite{Pachucki} gives the logs due to expanding the wave-function
at the origen $\sim (m\al)^3$ (which are naturally written in terms of
$\al(\nu)$) in terms of $\al_{\rm em}$. Therefore, we can trace back all
the logs of the computation in Ref.  \cite{Pachucki}. This provides
a check of our calculation to a level where it starts to first differ
with what would be the vNRQED result. Nevertheless, it may happen that, 
if the corrections of the vNRQCD results for the equal mass
calculation are finally confirmed, they may also explain the different
result obtained here.

\medskip

In conclusion, we have computed the energy spectrum at NNLL for an
Hydrogen-like system with $n_f$ massless fermions. We have checked our
results at $O(m\al^6\ln^2)$ by comparing with results already available
in the literature \cite{Pachucki} for muonic Hydrogen and
agreement has been found. We have also compared with what we would
expect to be the result in the vNRQED framework based on the rules of
\cite{MSS} and disagreement has been found. Finally, we would like to 
mention that the above results can be useful in checking higher order 
logs in computations of the spectrum for muonic atoms or alike where the 
electron can be considered to be a light particle.
   
\medskip

{\bf Acknowledgments}\\ 
We thank A. Hoang and specially J. Soto for discussions. We would also like 
to thank J. Soto for the reading of the manuscript.

\medskip

{\it Note added.}  After this paper was sent to the hep-ph archives a
new paper hep-ph/0204299 appeared in the net where it was pointed out
that there was a systematic error in the vNRQCD computations to date
and that the diagram Fig. 20b in this reference should be included in
such computations. After its inclusion, the corrected $1/m^2$
potential obtained in vNRQCD agrees with the $1/m^2$ potential
obtained in pNRQCD \cite{rgmass}, the replacement $c_D(\nu)
\rightarrow c_D(\nu^2/m)$ should be done in Eq. (\ref{anomdim}) and
the new result of vNRQED for the muonic hydrogen spectrum agrees with
our result.


\end{document}